# Collaborative Approaches to Enhancing Smart Vehicle Cybersecurity by AI-Driven Threat Detection

Syed Atif Ali
Cisco CCIE
Taxes, USA.
Syed.ali@technologyyours.com

Salwa Din
York University,ON
Toronto, Canada.
Email: salkammd@my.yorku.ca

## Abstract

The introduction sets the stage for exploring collaborative approaches to bolstering smart vehicle cybersecurity through AI-driven threat detection. As the automotive industry increasingly adopts connected and automated vehicles (CAVs), the need for robust cybersecurity measures becomes paramount. With the emergence of new vulnerabilities and security requirements, the integration of advanced technologies such as 5G networks, blockchain, and quantum computing presents promising avenues for enhancing CAV cybersecurity . Additionally, the roadmap for cybersecurity in autonomous vehicles emphasizes the importance of efficient intrusion detection systems and AI-based techniques, along with the integration of secure hardware, software stacks, and advanced threat intelligence to address cybersecurity challenges in future autonomous vehicles.

## 1. Introduction

The landscape of smart vehicle cybersecurity is continually evolving, with new vulnerabilities emerging alongside technological advancements. The Next Generation Mobile Networks Alliance has identified additional security requirements for 5G wireless networks, enabling real-time data offloading to cloud servers for cyberattack detection and mitigation. Additionally, the integration of 5G features supports Software-Defined Networking (SDN) and Network Function Virtualization (NFV), enhancing network scalability and management. Furthermore, blockchain technology presents a distributed and trust-based security solution, enabling connected and automated vehicles (CAVs) to establish trust among themselves, roadside infrastructure, and cloud servers. Quantum computing also holds promise for cryptography breakthroughs, with potential applications in secure key exchange and the development of quantum random number generators for automotive security [1].

In the context of autonomous vehicles, the need for efficient intrusion detection systems is emphasized, with a focus on state-of-the-art AI-based techniques. The roadmap for cybersecurity in autonomous vehicles underscores the importance of secure hardware and software stacks, coupled with advanced threat intelligence, as fundamental components in achieving various security goals [2]. These insights highlight the critical significance of collaborative efforts in academia, industry, and government to address the evolving cybersecurity challenges in smart vehicle technologies.

## 2. Smart Vehicle Cybersecurity

Smart vehicle cybersecurity encompasses a wide range of challenges and vulnerabilities that must be addressed to ensure the safety and security of smart vehicles. As highlighted by [1] , the emergence of new vulnerabilities necessitates the identification of additional security requirements, especially in the context of 5G wireless networks. The use of 5G networks enables edge devices to offload data into the cloud, facilitating cyberattack detection and mitigation. Additionally, technologies such as blockchain and quantum cryptography are being explored to enhance smart vehicle security and privacy, with a focus on V2X communication network and secure key exchange between ECUs in the in-vehicle network. Furthermore, [3] emphasize the various types of attacks that smart vehicles are vulnerable to, ranging from basic functionalities disruption to sophisticated network layer threats, such as DoS and DDoS attacks, Sybil attacks, spoofing, and GPS deception attacks. These threats necessitate the implementation of robust security measures across all layers to ensure the reliability and safety of vehicular communications.

These references provide insights into the multifaceted nature of smart vehicle cybersecurity, highlighting the need for collaborative and AI-driven approaches to address the evolving challenges and vulnerabilities in this domain.

### 2.1. Challenges and Vulnerabilities

In the realm of smart vehicle cybersecurity, there are numerous challenges and vulnerabilities that need to be addressed to ensure the safety and security of these vehicles. One of the key challenges is the potential emergence of new vulnerabilities, leading to the need for new security requirements [4]. For instance, the Next Generation Mobile Networks Alliance has identified additional security requirements for 5G wireless networks, which will impact the security of smart vehicles. Additionally, the interaction of connected autonomous vehicles (CAVs) with other smart systems in urban traffic poses a significant security concern, as it increases the exposure to potential attackers. Furthermore, the lack of a framework for connected autonomous systems hinders the design and implementation of secure autonomous vehicles, making it imperative to develop a security-by-design framework for AV from the first principle [1] [5] .

Moreover, the security of vehicle-to-everything (V2X) communication is crucial to prevent the spread of cyberattacks from CAVs to connected infrastructures and vice versa. Ensuring V2X security is essential given the concerns about the acquisition of sensitive information by attackers and the connection of CAVs with in-vehicle personal devices and the outside world. [6] Addressing these challenges and vulnerabilities requires a concerted effort from academia, industry, and government, emphasizing the urgency and necessity of adopting collaborative and AI-driven strategies to enhance cybersecurity in smart vehicles [7].

## 3. Artificial Intelligence in Cybersecurity

Artificial intelligence (AI) plays a pivotal role in enhancing cybersecurity measures for smart vehicles. With the advent of 5G networks, edge devices can leverage higher bandwidth to offload data into the cloud server in real time, facilitating cyberattack detection and mitigation [8] . Additionally, blockchain technology offers a decentralized and trust-based security solution that can be utilized by connected and automated vehicles (CAVs) to establish trust with each other, roadside infrastructure, and cloud servers. The distributed nature of blockchain makes it challenging for attackers to launch an attack, thereby enhancing CAV security and privacy. Furthermore, quantum computing holds the potential to create breakthroughs in AI, optimization, and cryptography, offering secure key exchange between electronic control units (ECUs) in the in-vehicle network [9].

AI algorithms have demonstrated superior performance in intrusion detection systems (IDS) and advanced driver-assistance systems (ADAS) for autonomous vehicles [2]. However, these algorithms are susceptible to carefully crafted adversarial attacks, and the increasing connectivity of vehicles may lead to common cyber threats such as Black-hole DDoS attacks and Sybil attacks. Moreover, the use of blockchain technology in autonomous vehicles can revolutionize future security measures by providing accurate and simultaneous access to different types of data, such as traffic information and vehicle tracking. As the landscape of smart vehicle cybersecurity continues to evolve, the integration of AI-driven threat detection and blockchain technology is becoming increasingly imperative to ensure robust security measures for smart vehicles.

## 3.1. Applications in Smart Vehicle Security

Artificial intelligence (AI) plays a pivotal role in fortifying the cybersecurity of smart vehicles. One key application is the development of efficient intrusion detection systems, which are essential for safeguarding smart vehicles against cyber threats [10]. [2] emphasize the significance of AI-based intrusion detection techniques as a crucial element of the cybersecurity roadmap for autonomous vehicles. These techniques, coupled with advanced threat intelligence, contribute to achieving various security goals and addressing open challenges in smart vehicle cybersecurity. Furthermore, AI techniques, such as transfer learning and machine learning-based misbehavior detection systems, have been instrumental in mitigating cyber-attacks within vehicular networks [3].

In addition to intrusion detection, AI is also leveraged for accurate traffic flow prediction, contributing to security enhancement in vehicular networks. These applications underscore the multifaceted role of AI in addressing specific security concerns and fortifying the defenses of smart vehicles against cyber threats.

# 4. Collaborative Approaches to Enhancing Smart Vehicle Cybersecurity

Collaborative approaches to enhancing smart vehicle cybersecurity are crucial in addressing the complex challenges posed by the cybersecurity of smart vehicles. Interdisciplinary collaboration is emphasized to leverage diverse expertise and perspectives in tackling these challenges effectively. [1] highlight the advantages of the 5G network, such as its higher bandwidth for real-time data offloading into cloud servers to facilitate cyberattack detection and mitigation. Additionally, the use of blockchain technology is proposed to establish trust among connected and automated vehicles (CAVs), roadside infrastructure, and cloud servers, thereby enhancing CAV security and privacy. Quantum cryptography, particularly quantum key distribution, is identified as a potential method to ensure secure key exchange between electronic control units (ECUs) in the in-vehicle network, addressing one of the biggest threats to in-vehicle security.

Furthermore, [2] stress the significance of tamper-proof AI algorithms in intrusion detection systems (IDS) and advanced driver-assistance systems (ADAS) for autonomous vehicles. The authors also underscore the increasing vulnerability to Black-hole DDoS attacks, Sybil attacks, and model inversion attacks, necessitating the creation of new approaches for tamper-proof and adversarial attack-resilient AI algorithms. Moreover, securing the supply chain of semiconductor integrated circuit components in vehicles is identified as crucial, especially with the increasing demand for roadside units (RSUs) and 5G infrastructure. Blockchain technology is proposed as a solution to provide accurate and simultaneous access to different types of data, mitigating security and privacy issues in autonomous vehicles. These insights underscore the need for collaborative efforts across academia, industry, and government to address the evolving cybersecurity challenges in smart vehicles.

### 4.1. Interdisciplinary Collaboration

Interdisciplinary collaboration plays a pivotal role in fortifying cybersecurity strategies for smart vehicles. Integrating insights and skills from diverse fields such as AI, cybersecurity, and automotive engineering can lead to more comprehensive and robust approaches to addressing emerging vulnerabilities and security requirements. For instance, the integration of 5G wireless networks can facilitate real-time data offloading to cloud servers, enabling efficient cyberattack detection and mitigation. Additionally, the use of blockchain technology establishes trust among connected autonomous vehicles (CAVs), roadside infrastructure, and cloud servers, thereby enhancing security and privacy through decentralized trusted V2X communication networks [1]. Furthermore, the development of state-of-the-art AI-based intrusion detection techniques and secure hardware and software stacks are identified as fundamental components in the cybersecurity roadmap for autonomous vehicles [2].

## 5. AI-Driven Threat Detection in Smart Vehicles

AI-driven threat detection in smart vehicles is a critical aspect of enhancing cybersecurity in the automotive industry. Leveraging the capabilities of AI, smart vehicles can proactively detect and mitigate potential threats, thereby contributing to overall cybersecurity posture. Techniques such as intrusion detection systems and state-of-the-art AI-based algorithms play a pivotal role in this domain, as highlighted by [2]. Furthermore, the integration of secure hardware and software stacks, along with advanced threat intelligence, is emphasized as vital components for achieving robust security goals in autonomous vehicles.

In addition to AI-driven threat detection, emerging technologies such as blockchain and quantum computing are also poised to revolutionize cybersecurity in smart vehicles [11]. [1] underscore the potential of blockchain in establishing trust among connected and automated vehicles (CAVs) and the roadside infrastructure, while also highlighting the significance of quantum cryptography for secure key exchange and quantum random number generators for automotive security. These advancements underscore the multidimensional approach required to fortify the cybersecurity framework of smart vehicles.

## 5.1. Techniques and Algorithms

In the realm of smart vehicles, AI-driven methods play a crucial role in enhancing cybersecurity through threat detection. The implementation of efficient intrusion detection systems is imperative, and state-of-the-art AI-based intrusion detection techniques have been identified as essential components in addressing cybersecurity concerns. According to [2], the integration of secure hardware and software stacks, along with advanced threat intelligence, is fundamental in achieving security goals for smart vehicles. This emphasizes the significance of AI-driven threat detection techniques in fortifying the cybersecurity of smart vehicles.

Moreover, [12] highlight the importance of sensor technology and computer vision in autonomous driving, which underscores the relevance of AI-driven algorithms for threat detection in smart vehicles. The comprehensive review of vehicle detection algorithms and their practical applications in autonomous driving demonstrates the critical role of AI-based techniques in the context of smart vehicles. By understanding these techniques, stakeholders can gain insights into the practical implementation of AI in enhancing cybersecurity for smart vehicles, as discussed in this subsection.